\documentclass[aps,twocolumn,showpacs]{revtex4}

\usepackage{amsmath}
\usepackage{amssymb}

\usepackage{graphicx}

\renewcommand{\exp}[1]{\ensuremath{{\mathrm e}^{#1}}}
\newcommand{\G}{\mathcal{G}}
\renewcommand{\Im}{\mathrm{Im}\,}

\begin{document}

\title{ Extraorinary Hall effect in hybrid ferromagnetic/superconductor
(F/S) bilayer}

\author{N. Ryzhanova$^{1}$, B. Dieny$^{2}$, C. Lacroix$^{3}$,
N. Strelkov$^{1}$,  A. Vedyayev$^{1,2}$}
\email{vedy@magn.phys.msu.su}
\address{
$^1$Department of Physics, M.~V.~Lomonosov Moscow State
University, 119899 Moscow, Russia \\
$^2$CEA/Grenoble, D\'epartement de Recherche Fondamentale sur la
Mati$\grave e$re Condens\'ee,
SP2M/NM, 38054 Grenoble, France \\
$^3$Laboratoire de Magn\'etisme Louis N\'eel, CNRS,
BP166, 38042 Grenoble, France} 

\begin{abstract}
\bigskip 

Extraordinary Hall effect (EHE) in bilayer F/S(N) was investigated
theoretically. The conductivity tensor $\sigma_{\alpha\beta}$ 
is calculated in the Kubo formalism with Green functions
found as the solutions of the Gorkov equations. We considered
diffuse transport in the ferromagnetic layer, taking into 
account as a main mechanism of electron resistivity s-d scattering.
In this model Gorkov equations for s-electrons in the ferromagnetic
layer remain linear and are solved easily. It is shown that 
Hall field $E^H$ for both F/S and F/N contacts are step-functions
of the coordinate perpendicular to the planes of the layers and have
zero value in S(N) layer. The Andreev reflection increases the value
of Hall constant $R_s$ for F/S case. The value of the Hall constant is
$R_H^{F/S} = R_H^{\rm bulk} 
(\sigma^{\uparrow} + \sigma^{\downarrow})^2 / 
4 \sigma^{\uparrow}\sigma^{\downarrow}$, 
where $\sigma^{\uparrow}$ and $\sigma^{\downarrow}$ are 
conductivities of electrons with up and down spins, and $R_H^{\rm bulk}$
is the Hall constant in the bulk ferromagnetic metal.
In fact, $R_H^{F/S}$ coincides with EHE constant of the bilayer
of two ferromagnetic metals with equal thickness 
and opposite directions of their magnetizations. 
So we can make a conclusion, that the ideal interface between 
ferromagnetic metal and superconductor may be considered 
like a mirror with inversion in spin space.  
\end{abstract}

\pacs{75.70.+a, 74.80.Dm, 73.40.-c, 72.20.My}

%  75.75.+a Magnetic properties of nanostructures
%  74.80.Dm Superconducting layer structures: superlattices, heterojunctions, 
%           and multilayers
%  73.40.-c Electronic transport in interface structures
%  72.20.My Galvanomagnetic and other magnetotransport effects

\maketitle

The perpendicular spin-dependent transport in hybrid 
ferromagnetic\-/superconductor (F/S) and F1/F2/S structures
has been previously investigated in Ref.\ \cite{deJong,Lambert,Ryzhanova2001}.
It was shown that the Andreev reflection~\cite{Andreev}
at F/S interface causes a mixing of up and down spin channels
and simultaneously a large spin accumulation on F/S interface
arises. The total influence of both effects increases the resistance of the 
system at $T=0$ comparing to its value when the superconductor
is in the normal state. In the case of F$^{\uparrow}$/F$^{\downarrow}$/S
these effects strongly influence the giant magnetoresistance (GMR)
of the F1/F2 bilayer. For certain values of the parameters, 
the GMR could be even completely suppressed \cite{Lambert,Ryzhanova2001}.

Another spin-dependent transport effect, which has not been 
considered so far in F/S sandwiches, is the extraordinary Hall effect 
(EHE)~\cite{Kondo,Giovannini}. The current $\vec{j}$ is assumed
to flow perpendicular to the interface, the magnetization $\vec{M}$ 
in the F-layer is in plane, and the Hall electrical field  
$\vec{E}^H$ arises in plane in the direction perpendicular
to $\vec{M}$. In such geometry, the Andreev reflection 
is expected to play a more complicated role than on the
current-perpendicular-to-plane (CPP) GMR because EHE combines
current-in-plane (CIP) and CPP features.

It is the purpose of this letter to investigate theoretically 
the EHE in F/S(N) bilayers in the diffuse regime in both
situations where the non-magnetic layer is in its superconducting 
or normal states. In the present model, we assume that conductivities 
for spin-up and spin-down channels are different and the contribution 
of F/S interface resistance to the total resistance is small.
This is the situation for which the CPP-GMR for F1/F2/S is destroyed 
by the Andreev reflection and spin accumulation
\cite{Lambert,Ryzhanova2001}.

\section{General definitions}

For geometry used in the model, the following relation proposed
by Smith and Sears~\cite{SmithSears} can be written
\begin{equation}
E_x^H=j_z(R_0H_y+R_sM_y),
\end{equation}
$R_0$ is the normal Hall effect coefficient whereas $R_s$
represents the EHE coefficient. $H_y$ and $M_y$ are 
respectively the external applied field and magnetization. The 
bias voltage is applied between the planes with
coordinates $z = -a$ and $z=b$, and $z = 0$ is the position of interface.

As it was shown in~\cite{Kondo,Giovannini} extraordinary Hall effect has
two different origins --- skew-scattering and side-jump.
As a first approach to EHE in these structures, we will investigate
only the first one, taking into account elastic scattering on impurities.
The approach that we use for calculation of the Hall coefficient for
layered system has been described in~\cite{Crepieux}. Following the same
method, we calculate the components of the conductivity
tensor using Kubo formula~\cite{Kubo}:
\begin{equation}
\label{sigma1}
\sigma_{\alpha\beta}=\frac{\hbar e^2}{\pi \Omega}
\mathrm{Tr}\left[ v_\alpha\left( G^+ - G^- \right)
v_\beta\left( G^+ - G^- \right) \right]
\end{equation}
which includes simple "bubble" diagonal conductivity and vertex
corrections for off-diagonal components, which are responsible
for the extraordinary Hall effect. In~(\ref{sigma1}) $v_\alpha$,
$v_\beta$ are velocity components, indices $\alpha$, $\beta$
take the values $x$, $y$, $z$, $G^{+(-)}$ are retarded and advanced 
Green functions, $\Omega$ is the volume of the system.

 For the system under consideration, which is homogeneous in $xy$-plane
and inhomogeneous in $z$ direction, it is convenient to use 
($\vec\kappa,z$)--representation, where $\hbar\vec\kappa$ is the in-plane
electron momentum. The Green function $G$ is defined by equations:
\begin{eqnarray}
G = G^0 + G^0 H^{\rm so} G^0
\nonumber \\
G^0 = G_{\rm eff} + G_{\rm eff} T G_{\rm eff}
\end{eqnarray}
where $G^0$ is the Green function of the system in the absence
of spin-orbit interaction, $G_{\rm eff}$ is the effective 
Green function, diagonal on in-plane vector $\kappa$, 
calculated in the coherent potential approximation (CPA) \cite{Soven}.
$T$ is a scattering matrix and $H^{so}$ is
the spin-orbit interaction.   
If to adopt for the ferromagnetic layer the model of totally 
disordered binary alloy A$_c$B$_{1-c}$ ($c$ is a concentration 
of the alloy's component A) we can write the expressions
for $T$-matrix in singe-site approximation and for 
matrix elements of $H^{\rm so}$ in explicit form:
\begin{eqnarray}
\lefteqn{T^{\sigma\,+}_{\varkappa\varkappa'}(z)=
\frac1N\sum_n\exp{i\vec\rho_n(\vec\varkappa-\vec\varkappa')}\times}
\label{T-matrix}\\
&  &  \times\frac{\delta^\sigma\nu(n,z)-{\Sigma^\sigma}^+}
{1-\left(\delta^\sigma\nu(n,z)- {\Sigma^{\sigma}}^+\right)
{\G^\sigma}^+(z,z)}\equiv\frac1N\sum_n
t^{\sigma\,+}_{\varkappa\varkappa'}(n,z) \nonumber\\
\lefteqn{H^{so}_{\varkappa\varkappa'}(z)=
\frac1N\sum_n\exp{i\vec\rho_n(\vec\varkappa-\vec\varkappa')}\times}
\label{Hso}\\
&  &  \nu(n,z)i\lambda M_y[\vec k\times\vec k']_y\equiv\frac1N\sum_n
H^{so}_{\varkappa\varkappa'}(n,z) \nonumber
\end{eqnarray}
In~(\ref{T-matrix}) and~(\ref{Hso}) $\delta^\sigma=\varepsilon^\sigma_A-
\varepsilon^\sigma_B$ is scattering parameter, $\varepsilon^\sigma_A$ and
$\varepsilon^\sigma_B$ are the band centers for A and B components depending
on spin $\sigma$, $\nu(n,z)$ is projection operator:
\begin{equation}
\nu(n,z)=a_B(n,z)c-a_A(n,z)(1-c)
\end{equation}
\begin{equation}
a_\alpha(n,z)=\left\{
\begin{array}{rl}
1,& \mbox{if the site (}\vec\rho_n,z\mbox{) is occupied by}\\
  & \alpha\mbox{ atom}\\
0,& \mbox{in the opposite case}
\end{array} \right.
\end{equation}
$\lambda=\lambda_A-\lambda_B$ is parameter of spin-orbit scattering
which is non-zero only in the ferromagnetic layer. ${\Sigma^\sigma}^+$
is the coherent potential which is the solution of self-consistent
equation $\left<{T^\sigma}^+(z)\right>=0$, where $\langle\dots\rangle$ means
averaging on impurities distribution, ${\G^\sigma}^+(z,z)=
\frac1N\sum_\varkappa G^{\sigma\, +}_{\vec\varkappa}(z,z)$.
In~(\ref{Hso}) the $z$-component of electron momentum vector $\vec k$ in
$[\vec k\times\vec k']_y$ is the antisymmetric gradient operator:
$k_z=i \left(\stackrel\to\nabla_z-\stackrel\gets\nabla_z\right)$.

It is important to note that from definitions~(\ref{T-matrix})
and~(\ref{Hso}) the first non-zero contribution into the vertex correction of
formula~(\ref{sigma1}) linear on $H^{so}$ is that containing
$\left< t^{\sigma\, +}_{\varkappa\varkappa'}(n,z)
H^{so}_{\varkappa\varkappa'}(n',z')\right>\sim\delta_{nn'}\delta(z-z')$.

In the adopted geometry the system of equations for Hall fields
can be written as follows:
\begin{subequations}
\begin{align}
j_x(z)=&\int\sigma_{xx}(z,z')E_x^H(z')\,dz'+\nonumber\\
&+\iint\sigma_{xz}(z,z',z'')E_z(z')\,dz'\,dz''=0 \label{jx}\\
j_z(z)=&\int\sigma_{zz}(z,z')E_z(z')\,dz'+\nonumber\\
&+\iint\sigma_{zx}(z,z',z'')E_x^H(z')\,dz'\,dz'' \label{jz}
\end{align}
\end{subequations}
We consider $H^{\rm so}/\delta$ like a small parameter of the theory.
So $\sigma_{zx} \ll \sigma_{zz}$ and the second term
in equation (\ref{jz}) can be omitted. 
The off-diagonal component of conductivity has a three-point character. 
The additional coordinate $z''$ represents the scattering plane.

\section{Model}

We consider a bilayer of the type F/S, where F is ferromagnetic layer, 
S is a superconducting layer. A simple two band (spin up and down) free 
electron model is adopted for this calculation. 
The Hamiltonian of the system is therefore written as:
\begin{subequations}
\begin{equation}
H=H_F+H_S \label{H}
\end{equation}
\begin{multline}
H_F=\sum_{\sigma=+(\downarrow),-(\uparrow)}
\;\int\limits_{r\in F}\left[\left(\frac{\hat p^2}{2m}-
\varepsilon_F+\mathrm{sign}(\sigma)
\varepsilon_{\mathrm{ex}}\right)\right.\\
\times\psi_\sigma^{s*}(r)\psi^{s}_\sigma(r)
+\gamma_{sd}(r)\left(\psi_\sigma^{s*}(r)
\psi_\sigma^d(r)+h.c.\right)\biggr]\,d^3r\label{Hf}
\end{multline}
\begin{multline}
H_S=\int\limits_{r\in S}\left[\sum_\sigma\left(\frac{\hat
p^2}{2m}-\varepsilon_F\right) \psi_\sigma^{s*}(r)
\psi_\sigma^{s}(r)\right. \\
+\left(\Delta(r)\psi_\uparrow^{s*}(r)\psi_\uparrow^{s*}
(r)+h.c.\right)\biggr]\,d^3r\label{Hs}
\end{multline}
\end{subequations}
where $\displaystyle\varepsilon_{\mathrm{ex}}=\frac{p^{\uparrow2}_F -
p^{\downarrow2}_F}{2m}$ is the exchange energy, $\varepsilon_F$,
$p^\sigma_F$ are respectively the Fermi energy and momentum.
The second term in~(\ref{Hf}) describes the scattering of quasi free 
s-electrons into almost localized d-states. In bulk
ferromagnetic metals d-states may give contribute to the
current~\cite{Brouers}. However in the present situation, 
we consider that there are no d-states in the superconductor. 
Therefore d-electrons are completely 
reflected on F/S interface and do not contribute to the current. 
On other respects, s-d scattering in ferromagnetic 
dirty d-metal alloys remains the most important mechanism 
of s-electrons scattering \cite{Brouers}. 
In the case under consideration,
we take it into account and consider that the random s-d scattering potential 
$\gamma_{sd}$ is much smaller then $\varepsilon_F$. We further 
calculate the mean free path in the Born approximation. $\Delta$
is the order parameter in the superconductor. Now the system 
of Gorkov equations \cite{Gorkov} for the normal $G_{\rm eff}$ and 
anomalous $F_{\rm eff}$ Green functions can be written:
\begin{subequations}
\label{Gorkov}
\begin{multline}
\left[\frac{\hbar^2}{2m}\left(\frac{\partial^2}{\partial z^2}-
\varkappa^2\right)+\varepsilon_F+\varepsilon_{\mathrm{ex}}-
\gamma_{sd}^2G_{dd}^{\uparrow\uparrow}(z,z)\right]\\
\times G_{ss}^{\uparrow\uparrow}(z,z')
+\Delta F_{ss}^{\downarrow\uparrow}(z,z')=\delta(z-z')\label{Gorkov1}
\end{multline}
\begin{multline}
\Delta^*G_{ss}^{\uparrow\uparrow}(z,z')-
\left[\frac{\hbar^2}{2m}\left(\frac{\partial^2}{\partial z^2}-
\varkappa^2\right)+\varepsilon_F
-\varepsilon_{\mathrm{ex}}\right.\\
\left.+\gamma_{sd}^2 G_{dd}^{\downarrow\downarrow}(z,z)\right
]F_{ss}^{\downarrow\uparrow}(z,z')=0\label{Gorkov2}
\end{multline}
\end{subequations}
We ommited index "eff" for the brevity. The terms $\varepsilon_{\mathrm{ex}}$
and $\gamma^2G_{dd}^{\uparrow\uparrow(\downarrow\downarrow)}$ are different from zero
and $\Delta=0$ if $z$ belongs to F-layer and vice versa in 
S-layers. The system~(\ref{Gorkov})
is written for spin $\uparrow$. For spin $\downarrow$,
$\varepsilon_{\mathrm{ex}}$ has to be changed to $-
\varepsilon_{\mathrm{ex}}$ and
$G_{dd}^{\downarrow\downarrow(\uparrow\uparrow)}$ to
$G_{dd}^{\uparrow\uparrow(\downarrow\downarrow)}$. The main
difference between system~(\ref{Gorkov}) and usually employed
equation for F/S structures (see for example~\cite{Bulaevskii}) is that 
we took into account s-d scattering. 
Moreover, we considered this s-d scattering as the main mechanism
determining the mean free path of $s$-electrons~\cite{Brouers}.
The function $G_{dd}^{\sigma\sigma}(z,z)$ may be
considered like a constant in $z$-space if it is averaged over
the short wave length $(\hbar/p_F)$ oscillations and the system of
equation~(\ref{Gorkov}) may be solved analytically. Further,
we set $\varepsilon_{\mathrm{ex}}=0$ for s-electron. The explicit
expression for Green functions are:
\begin{subequations}
\begin{gather}
\G^{\uparrow\uparrow\, +}_{11}(-\infty<z,z'<0)=
\displaystyle
\frac{\exp{ik_1|z-z'|}}{2ik_1}\\
F^{\downarrow\uparrow\, +}_{11}(-\infty<z,z'<0)=
\displaystyle
\frac{\exp{ik_2^*z}\exp{-ik_1z'}}{k_1+k_2^*}\\
G^{\uparrow\uparrow\, +}_{12}(-\infty<z<0<z'<\infty)=
\displaystyle
\frac{\exp{-ik_1^*z}\exp{ik_sz'}}{i(k_1+k_s)}\\
F^{\downarrow\uparrow\, +}_{12}(-\infty<z<0<z'<\infty)=
\displaystyle
\frac{\exp{ik_2^*z}\exp{-ik_s^*z'}}{k_s^*+k_2^*}
\end{gather}
\begin{multline}
G^{\uparrow\uparrow\, +}_{22}(0<z,z'<\infty)=\\
\frac1{4ik_s}\left[\exp{ik_sz}\exp{-ik_s^*z'}+\exp{ik_s|z-z'|}\right]+\\
+\frac1{4ik_s^*}\left[\exp{ik_sz'}\exp{-ik_s^*z}-\exp{-ik_s^*|z-z'|}\right]
\end{multline}
\begin{multline}
F^{\downarrow\uparrow\, +}_{22}(0<z,z'<\infty)=\\
\frac1{4ik_s}\left[\exp{ik_sz}\exp{-ik_s^*z'}+\exp{ik_s|z-z'|}\right]-\\
-\frac1{4ik_s^*}\left[\exp{ik_sz'}\exp{-ik_s^*z}-\exp{-ik_s^*|z-z'|}\right]
\end{multline}
\end{subequations}
where $k_{1,2}=\sqrt{k_F^2-\varkappa^2+i\frac{2k_F}{l_{1,2}}}\equiv
c_{1,2}+id_{1,2}$; $k_s=\sqrt{k_F^2-\varkappa^2+i|\Delta^2|}\equiv
c_s+id_s$, $l_{1,2}$ are mean free paths for up and down spins.

Perpendicular transport in bilayer F/S was investigated
in~\cite{Ryzhanova2001} so we remind here only the result:
$E_1^\uparrow l_1=E_1^\downarrow l_2$, $E_1^\uparrow$ and
$E_1^\downarrow$ are effective electrical fields acting on
the carriers with spin up, down, $E_1^\uparrow=
\frac{2Vl_2}{a(l_1+l_2)+\frac43l_1l_2}$, $j_z\sim E_1^\uparrow l_1$.
Off-diagonal components of conductivity in~(\ref{sigma1}) as
well as diagonal one has two contributions -- normal and anomalous.
For example, one has the form:
\begin{multline}
\label{sigmanorm}
[\sigma_{xz}^{\mathrm{norm}}(z,z',z'')]_\varkappa\sim
L\Bigl<v_{x\varkappa} G^+_\varkappa(z,z'')\\
\left(T^+(z'') + H^{so}(z'')\right)_{\varkappa\varkappa'}
G^+_{\varkappa'}(z'',z')v_{z\varkappa'}
G^-_{\varkappa'}(z',z'')\\
\left( T^-(z'') + H^{so}(z'')\right)_{\varkappa'\varkappa}
G^-_\varkappa(z'',z)
\Bigr>
\end{multline}
where $L$ means linear on $H^{so}$ part of this expression.
Since $T^+_{\varkappa\varkappa'}
= T^+_{\varkappa'\varkappa}$ and $H^{so}_{\varkappa\varkappa'}=
- H^{so}_{\varkappa'\varkappa}$, term~(\ref{sigmanorm}) is
proportional to $\left<2iH^{so}(z)\Im T^+(z)\right>$ as
well as all other contributions to $\sigma_{xz}$. In the Born
approximation, we can write down for disordered binary system:
\begin{multline}
\Bigl<2iH^{so}(z)\Im T^+(z)\Bigr>=
-\lambda M_y c(1-c)(1-2c)
\left(\delta^\sigma\right)^2\times\\
\times \Im {\G^{\sigma\, +}}(z,z)
[\vec k\times\vec k']_y
\end{multline}
from which it is easy to see that this term is proportional 
to $(1-2c)/l_\sigma$.

 The unknown Hall field has to be found as a solution of the 
integral equation~(\ref{jx}). To solve it we take as a probe 
function for $E^H(z)$ the step function, taking  
value $E_1^{H\uparrow(\downarrow)}$ inside the ferromagnetic
layer and $E_s^{H\uparrow(\downarrow)}$ in the superconductor
for each direction of the electron's spin $\uparrow$, $\downarrow$.
In this case the equation (\ref{jz}) may be 
rewritten as a system of two equations: 
\begin{subequations}
\begin{multline}
E^{H\,\uparrow}_1l_1\int\frac{\varkappa^3}{k_F^3c}
\left(1-\frac{\exp{-2d_1(z+a)}}2-\frac{\exp{2d_1z}}2\right)\,d\varkappa+\\
+1/2E^{H\,\downarrow}_1l_2\int\frac{\varkappa^3}{k_F^3c}
\exp{2d_1z}\left(1-\exp{-2d_2a}\right)\,d\varkappa+\\
+1/2\left(E^{H\,\uparrow}_s+E^{H\,\downarrow}_s\right)
\int\frac{\varkappa^3}{k_F^2c^2d_s}
\exp{2d_1z}\left(1-\exp{-2d_sb}\right)\,d\varkappa=\\
=R_s^{\mathrm{bulk}}E_1^\uparrow l_1M_y
\frac{\sigma^{\uparrow}+\sigma^{\downarrow}}4
\int\frac{\varkappa^3}{k_F^3c}\Bigl(1-\\
\frac{\exp{-2d_1(z+a)}}2-
\frac{\exp{2d_1z}\exp{-2d_2a}}2\Bigr)\,d\varkappa\mbox{;}-a<z<0
\label{eq1}
\end{multline}
\begin{multline}
E^{H\,\uparrow}_1l_1\int\frac{\varkappa^3}{k_F^3c}
\exp{-2d_sz}\left(1-\exp{-2d_1a}\right)\,d\varkappa+\\
+E^{H\,\downarrow}_1l_2\int\frac{\varkappa^3}{k_F^3c}
\exp{-2d_sz}\left(1-\exp{-2d_2a}\right)\,d\varkappa+\\
+\left(E^{H\,\uparrow}_s+E^{H\,\downarrow}_s\right)
\int\frac{\varkappa^3}{k_F^2c^2d_s}\Bigl(1-\\
\frac{\exp{-2d_s(z+b)}}2-\frac{\exp{2d_s(z-b)}}2\Bigr)\,d\varkappa=\\
=R_s^{\mathrm{bulk}}E_1^\uparrow l_1M_y
\frac{\sigma^{\uparrow}+\sigma^{\downarrow}}2
\int\frac{\varkappa^3}{k_F^3c}\exp{-2d_sz}\Bigl(1-\\
\frac{\exp{-2d_1a}}2-\frac{\exp{2d_2a}}2\Bigr)\,d\varkappa;\quad
0<z<b
\label{eq2}
\end{multline}
\end{subequations}
where $R_s^{\mathrm{bulk}}$ is the Hall coefficient 
and $\sigma^{\uparrow(\downarrow)}$ are conductivities 
of up and down spin channels for the bulk ferromagnet.
For spin down we have to change 
$\uparrow\leftrightarrows\downarrow$; $1\leftrightarrows2$. 

 Solving the system of equations (\ref{eq1}), (\ref{eq2}) we found
that indeed the solution for the Hall field $E^H(z)$
in~the form of the step-like function satisfies the system 
(\ref{eq1}), (\ref{eq2}) and consequently the integral 
equation (\ref{jx}) for any~$z$. The results of the Hall
fields are: 
$ E_s^{H\,\uparrow} = E_s^{H\,\downarrow} =~0$, 
$E^{H\,\uparrow}_1l_1=E^{H\,\downarrow}_1l_2$,
$E^{H\,\uparrow}_1 +  E^{H\,\downarrow}_1 = R_s^{\mathrm{bulk}}
E^{\uparrow}_1 M_y \frac{\displaystyle \sigma^{\uparrow} + \sigma^{\downarrow}}
{\displaystyle 2}$.
So the extraordinary Hall coefficient for F/S bilayer is:
\begin{equation}
\label{RsFS}
R_s^{F/S}=\frac{(\sigma^{\uparrow} + \sigma^{\downarrow})^2}
{4 \sigma^{\uparrow} \sigma^{\downarrow}}R_s^{\mathrm{bulk}}
\end{equation}

\section{F/N bilayer}

Now we recalculate the Hall coefficient when the non-magnetic
layer is in the normal state. In this case, all Green functions 
are diagonal in the spin space:
\begin{subequations}
\begin{gather}
G^{\uparrow\uparrow\, +}_{11}(-\infty<z,z'<0)=
\displaystyle\frac{\exp{ik_1|z-z'|}}{2ik_1}\\
G^{\uparrow\uparrow\, +}_{12}(-\infty<z<0<z'<\infty)=
\displaystyle\frac{\exp{-ik_1^*z}\exp{ik_3z'}}{i(k_1+k_3)}\\
G^{\uparrow\uparrow\, +}_{22}(0<z,z'<\infty)=
\displaystyle\frac{\exp{ik_3|z-z'|}}{2ik_3}
\end{gather}
\end{subequations}
where $\displaystyle k_3=\sqrt{k_F^2-\varkappa^2+i\frac{2k_F}{l_3}}\equiv
c_3+id_3$, $l_3$ is the mean free path in the normal metal.

The following expressions are then obtained:
\begin{subequations}
\begin{equation}
E_1^\downarrow=E_1^\uparrow\frac{al_3+bl_1+ \frac{4}{3}l_1l_3}
{al_3+bl_2+ \frac{4}{3} l_2l_3}
\end{equation}
\begin{equation}
E_1^\uparrow=\frac{Vl_3}{2(al_3+bl_1+ \frac{4}{3} l_1l_3)}
\end{equation}
\end{subequations}

The system of equations for the Hall fields assumed to be 
step-functions can be written in the form:
\begin{subequations}
\begin{multline}
E^{H\,\uparrow}_1l_1\int\frac{\varkappa^3}{k_F^3c}
\left(1-\frac{\exp{-2d_1(z+a)}}2-\frac{\exp{2d_1z}}2\right)\,d\varkappa+\\
+1/2E^{H\,\uparrow}_2l_3\int\frac{\varkappa^3}{k_F^3c}
\exp{2d_1z}\left(1-\exp{-2d_3b}\right)\,d\varkappa=\\
=R_s^{\mathrm{bulk}}E_1^\uparrow l_1M_y
\frac{ \sigma^{\uparrow} + \sigma^{\downarrow} }4
\int\frac{\varkappa^3}{k_F^3c}\times\\
\times\Bigl(1-\frac{\exp{-2d_1(z+a)}}2-
\frac{\exp{2d_1z}}2\Bigr)\,d\varkappa;\ -a<z<0
\end{multline}
\begin{multline}
1/2E^{H\,\uparrow}_1l_1\int\frac{\varkappa^3}{k_F^3c}
\exp{-2d_3z}\left(1-\exp{-2d_1a}\right)\,d\varkappa+\\
+E^{H\,\uparrow}_2l_3\int\frac{\varkappa^3}{k_F^3c}
\left(1-\frac{\exp{-2d_3z}}2-\frac{\exp{2d_3(z-b)}}2\right)\,d\varkappa=\\
=R_s^{\mathrm{bulk}}E_1^\uparrow l_1M_y
\frac{\sigma^{\uparrow} + \sigma^{\downarrow}}4
\int\frac{\varkappa^3}{k_F^3c}\exp{-2d_3z}\times\\
\times\Bigl(1-\exp{-2d_1a}\Bigr)\,d\varkappa;\ 0<z<b
\end{multline}
\end{subequations}
For spin down we have to change $\uparrow\leftrightarrows\downarrow$;
$1\leftrightarrows2$.
Solution of this system gives us
$E^{H\,\uparrow}_2=E^{H\,\downarrow}_2=0$,
\begin{subequations}
\begin{equation}
E^{H\,\uparrow}_1=
\frac{Vl_3( \sigma^{\uparrow} + \sigma^{\downarrow} )M_y}
{4(al_3+bl_1+ \frac{4}{3} l_1l_3)}
R_s^{\mathrm{bulk}}
\end{equation}
\begin{equation}
E^{H\,\downarrow}_1=
\frac{Vl_3( \sigma^{\uparrow} + \sigma^{\downarrow})M_y}
{4(al_3+bl_2+ \frac{4}{3} l_2l_3)}
R_s^{\mathrm{bulk}}
\end{equation}
\end{subequations}
where $V$ is the total voltage drop across the F/N bilayer, 
and Hall coefficient is:
\begin{equation}
\label{RsFN}
R_s^{F/N}=\frac{(l_1+l_2)}2
\frac{2al_3+b(l_1+l_2)+ \frac{4}{3} l_3(l_1+l_2)}
{al_3(l_1+l_2)+2bl_1l_2+ \frac{8}{3} l_1l_2l_3}
R_s^{\mathrm{bulk}}.
\end{equation}

With the same method it is easy to show that the Hall 
constant for a spin-valve bilayer in antiparallel magnetic 
configuration F$^\uparrow$/F$^\downarrow$ is equal to 
$\displaystyle R_s^{F_1/F_2}=
\frac{(  \sigma^{\uparrow} + \sigma^{\downarrow} )^2}
{4 \sigma^{\uparrow}\sigma^{\downarrow} }R_s^{\mathrm{bulk}}$.

\section{Conclusions}

As can be see from~(\ref{RsFS}) and~(\ref{RsFN}), the relative
change of the Hall coefficient in the presence of superconducting
contact is equal:
\begin{equation}
\frac{R_s^{F/S}-R_s^{\mathrm{bulk}}}{R_s^{\mathrm{bulk}}}=
\frac{(\sigma_{\uparrow} - \sigma_{\downarrow})^2}
{4 \sigma_{\uparrow} \sigma_{\downarrow} }\nonumber
\end{equation}
and
\begin{equation}
R_s^{F/S}-R_s^{F/N}\sim( \sigma_{\uparrow} - \sigma_{\downarrow} )^2\nonumber
\end{equation}
Simultaneously it is shown that resistivity $\rho^{F/S}$ of F/S bilayer
is equal $\displaystyle \rho^{F/S}=\frac{\rho_1+\rho_2}2$, 
where $\rho_1$ and $\rho_2$ are resistivities of up and down spin 
channels, this resistivity as well as $R_s^{F/S}$ coincides 
correspondingly with resistivity and the Hall
constant of the bilayer of two ferromagnetic metals with opposite
direction of magnetizations. Therefore, we conclude that
an ideal interface between a ferromagnetic metal and a superconductor
can be considered like a quantum mirror with inversion in spin-space.
Of course, the roughness of the interface may spoil the mirror image.
In addition we considered the case where the spin-diffusion 
length was much larger than the thickness of the ferromagnetic layer. 
The influence of spin-flip processes on Hall 
effect will be considered in a forth coming paper.

Experimental investigation of EHE in the situation, close to one
described in the letter, may be done if to use as a ferromagnetic layer
the alloy  CuNi with relatively small exchange splitting and 
high resistance. The thickness  $t^{\rm CuNi}$
of CuNi layer has to be in the interval
$l_{\rm el} \ll t^{\rm CuNi} \ll l_{\rm sd}$, where
$l_{\rm sd}$ is the spin-diffusion length
and $l_{\rm el}$ is the longest mean free path 
of the electron in bulk CuNi.

\section*{Acknowledgments}

N.R.\ acknowledges the Louis N\'eel Laboratory (CNRS/\-Grenoble, France)
and A.V.\ acknowledges CEA/\-Grenoble\-/DRFMC/\-SP2M/NM for hospitality.
This work was supported by Russian Foundation for Basic Research.

\end{document}